\documentclass[
 twocolumn,
 superscriptaddress,
 bibnotes,
 amsmath,amssymb,
 prf,
 longbibliography
]{revtex4-2}

\usepackage{graphicx}
\usepackage{color}
\usepackage{dcolumn}
\usepackage{bm}
\usepackage{hyperref}
\usepackage{bigints}
\usepackage[dvipsnames]{xcolor}

\usepackage{textcomp}

\usepackage{rotating}

\usepackage{xcolor}
\usepackage{newfloat}

\begin{document}

\title{Droplets sliding on soft solids shed elastocapillary rails}

\author{Nan~Xue}
\affiliation{Department of Materials, ETH Z\"urich, 8093 Z\"urich, Switzerland.}
\affiliation{Department of Materials Science and Engineering, Cornell University, Ithaca, NY, 14853, USA.}
\affiliation{Laboratory of Atomic and Solid-State Physics, Cornell University, Ithaca, NY, 14853, USA.}
\author{Lawrence~A.~Wilen}
\affiliation{Center for Engineering Innovation and Design, School of Engineering and Applied Sciences, Yale University, NH, 06520, USA}
\author{Robert~W.~Style}
\email{robert.style@mat.ethz.ch}
\affiliation{Department of Materials, ETH Z\"urich, 8093 Z\"urich, Switzerland.}
\author{Eric~R.~Dufresne}
\email{eric.r.dufresne@cornell.edu}
\affiliation{Department of Materials, ETH Z\"urich, 8093 Z\"urich, Switzerland.}
\affiliation{Department of Materials Science and Engineering, Cornell University, Ithaca, NY, 14853, USA.}
\affiliation{Laboratory of Atomic and Solid-State Physics, Cornell University, Ithaca, NY, 14853, USA.}

\date{\today}

\begin{abstract}
The surface tension of partially wetting droplets deforms soft substrates.
These deformations are usually localized to a narrow region near the contact line, forming a so-called `elastocapillary ridge.'
When a droplet slides along a substrate, the movement of the elastocapillary ridge dissipates energy in the substrate and slows the droplet down.
Previous studies have analyzed isotropically spreading droplets and found that the advancing contact line `surfs' the elastocapillary ridge, with a velocity determined by a local balance of capillary forces and bulk rheology.
Here, we experimentally explore the dynamics of a droplet sliding across soft substrates.
At low velocities, the contact line is nearly circular, and dissipation increases logarithmically with speed.
At higher droplet velocities, the contact line adopts a bullet-like shape, and the dissipation levels off.
At the same time,  droplets shed a pair of `elastocapillary rails' that fade away slowly behind it.
These results suggest that droplets favor sliding along a stationary ridge over surfing atop a translating one.
\end{abstract}

\maketitle

\section{Introduction}

A liquid droplet placed on a soft substrate induces the formation of an elastocapillary wetting ridge along its contact line \cite{shanahan1994anomalous, shanahan1995viscoelastic, pericet2008effect, jagota2012surface, style2017elastocapillarity, bardall2018deformation, andreotti2020statics}.
These ridges arise from the interplay between capillary forces and the substrate's elastic response. 
They are localized within a narrow region with a characteristic height of $\gamma / E$, where $\gamma$ represents the surface tension of the droplet and $E$  Young's modulus of the soft substrate.
For example, a water droplet on a gel with an elastic modulus of ${O}(10)$ kPa will typically form ridges with heights of ${O}(10)$ $\mu$m.

Previous studies \cite{carre1996viscoelastic, long1996static, karpitschka2015droplets, zhao2018geometrical, van2018dynamic, liang2018surface, van2020spreading, smith2021droplets, mokbel2022stick, jeon2023moving} have shown that formation of an elastocapillary ridge underneath a moving contact line introduces dissipation within the substrate,  slowing it down.
This phenomenon is called `viscoelastic braking'.
Two-dimensional models integrating capillarity and viscoelasticity have been successful in quantifying the dissipation associated with viscoelastic braking for isotropically spreading droplets \cite{karpitschka2015droplets, zhao2018geometrical, van2018dynamic, van2020spreading, smith2021droplets, mokbel2022stick, jeon2023moving}.
Essentially, ridge formation and relaxation generate a drag force per unit length of the contact line that increases with velocity, up to a limiting value comparable to the surface tension.
Beyond this point, the contact line detaches from the elastocapillary ridge, leading to stick-slip motion.

For sliding droplets, the situation is more complex.
Directional shear stresses that drive sliding break symmetry and deform the droplet \cite{podgorski2001corners, oleron2022dynamic, oleron2024morphology}.
Now, we expect feedback between the contact line dissipation and the global shape of the droplet. 
This competition plays out across a wide range of length scales, from the micron-scale structure of the elastocapillary ridge, to the millimeter-scale shape of the droplet.

In this paper, we apply interferometry to measure sub-micron deformations across millimeter-scale sliding droplets while simultaneously quantifying global dissipation, through drag force.
At low velocities, the contact line is nearly circular, and drag increases logarithmically with speed.
At higher velocities, the drag force plateaus as the contact line becomes bullet-shaped.
Strikingly, we also observe a pair of parallel ridges, which we call `elastocapillary rails,' that gradually fade behind the droplet.
The formation of elastocapillary rails is indicative of feedback between the microscopic elastocapillary ridge and the macroscopic droplet shape.

\begin{figure*}[ht]
 \centering
 \includegraphics[width=1\textwidth]{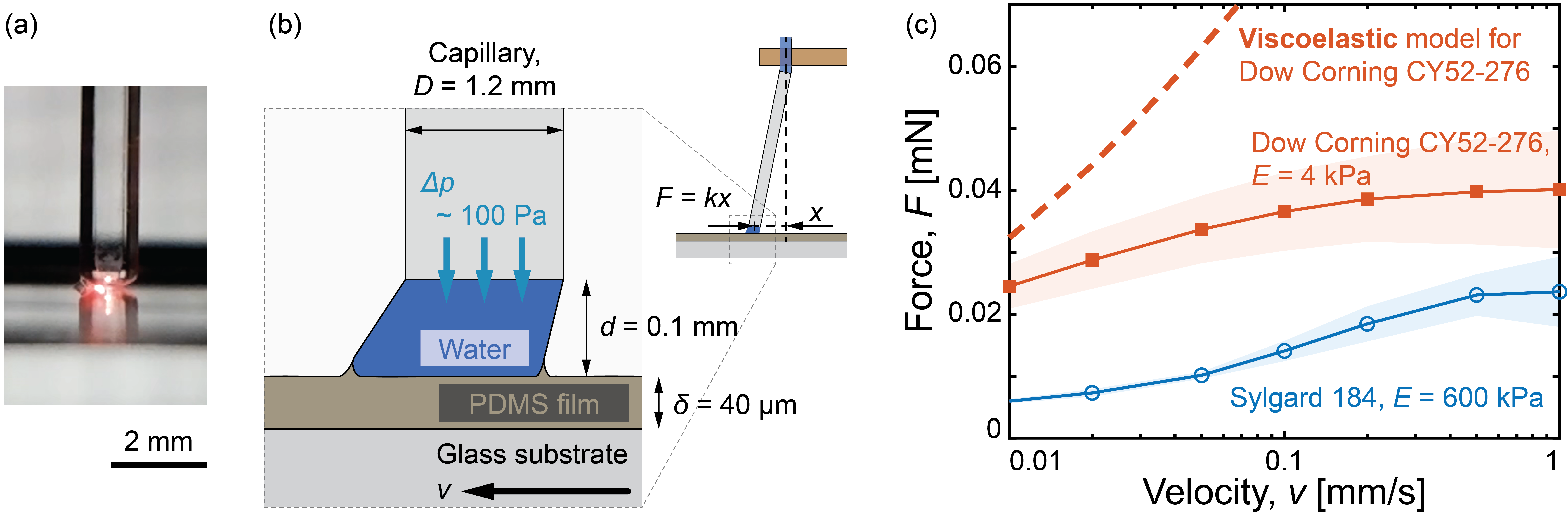}
 \caption{The platform for studying droplet motion on soft substrates.
 (a) Side view of the droplet between the capillary and the substrate; see also in Movie 1.
 (b) Schematic of the experimental setup.
 (c) The measured force $F$ as a function of the droplet velocity $v$.
 {The dashed line represents the prediction of a 2D, linear, viscoelastic model for a moving, circular contact line, with a diameter equal to that of the resting droplet.}
 The colored shadow represents the standard deviation across three experiments on different samples.}
 \label{fig:1}
\end{figure*}

\section{Experimental Methods}

We employ an interferometric platform \cite{xu2020viscoelastic} to simultaneously investigate the interplay between macroscopic droplet shape, microscopic elastocapillary ridge formation and relaxation, and the associated energy dissipation during droplet motion on soft substrates.
Figure~\ref{fig:1}(a) and Movie 1 depict the side view of the platform, while Fig.~\ref{fig:1}(b) provides a schematic representation (optical paths detailed in Ref. \citenum{xu2020viscoelastic} and {Fig.~S1} in ESI).

\subsection{Experimental setup}

An aqueous liquid bridge is formed between a capillary (outer diameter $D = 1.2$ mm) and a soft substrate [Fig.~\ref{fig:1}(b)].
We refer to this as a `droplet' for the rest of the paper.
The capillary connects to a water reservoir through silicone rubber tubing, allowing us to maintain a constant positive pressure difference ($\Delta p = 100$ Pa) between the interior of the droplet and the surrounding atmosphere.
This compensates for the water evaporation, enabling experiments using pure water instead of ionic liquids \cite{khattak2022direct}.

Two silicone materials with differing stiffnesses are used as substrates:  a stiffer elastomer (Sylgard 184, 10:1, $E = 600$ kPa) and a softer gel (Dow Corning CY52-276, 1:1, $E = 4$ kPa) \cite{xu2020viscoelastic}.
Both exhibit similar macroscopic wetting properties with a surface tension $\gamma_\mathrm{s} \approx 30$ mN/m \cite{xu2020viscoelastic}.
Following Refs.~\citenum{karpitschka2015droplets, xu2020viscoelastic}, the complex modulus of the softer gel is modeled by the power-law form $G^*(\omega) = G_0 [1 + (i \omega \tau )^n]$ with shear modulus $G_0 = 1.3$ kPa, intrinsic timescale $\tau = 0.11$ s, and power $n = 0.54$.
We prepare uniform 40 $\mu$m thick gel films by spin-coating uncured silicone (five drops from a 3 mL pipette, one minute at 800 rpm) onto glass substrates, followed by curing at 40°C over {25 hours}.
In our experiments, we observed no discernible effects attributable to a potentially uncured gel \cite{hourlier2017role}.

The droplet height is set at $d = 0.1$ mm by controlling the capillary-substrate distance.
Motorized actuators (Thorlabs, Z812B and ZST213B) control the lateral positions of the capillary and substrate, {respectively}, each enabling relative velocities between $v = 0.01$ mm/s and $v = 1$ mm/s.
For high-accuracy dissipation measurements, the base of the capillary is stationary while the substrate moves at the velocity, $v$.
Conversely, to measure substrate deformation via interferometry, the substrate is stationary, and the base of the capillary moves at the velocity, $v$.
This setup allows for precise control of the droplet's velocity and measurement of its response.
Throughout this manuscript, we report data only for droplets in a steady state, ensuring they have been translated at least 2 mm from their initial position.

\subsection{Force and Deformation Measurements}

Similar to Refs.~\citenum{khattak2022direct, hinduja2024slide}, the displacement of the capillary tip ($x$) responds linearly to the force ($F$) acting on the sliding droplet [Fig.~\ref{fig:1}(b)].
{To calibrate the spring constant, $k$, we re-orient the capillary horizontally, bring its tip in contact with a balance, and record the force response to vertical displacements (details in {Fig.~S2} in ESI).}
By adjusting the capillary and tubing geometry, we maintain $k \approx 0.1$ mN/mm throughout our experiments.
Figure~\ref{fig:1}(c) presents typical force measurements as a function of droplet velocity.

\begin{figure*}[ht]
 \centering
 \includegraphics[width=1\textwidth]{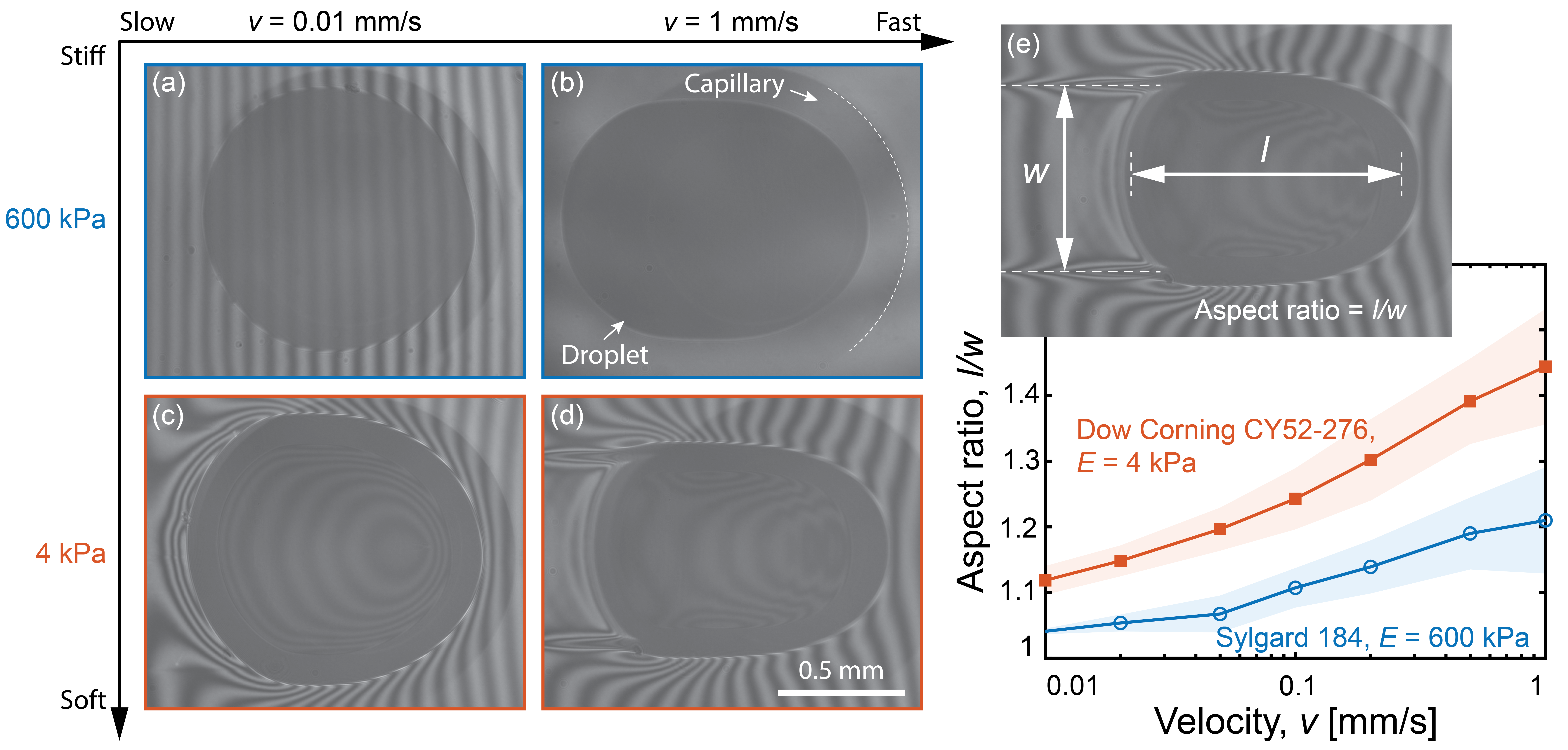}
 
 \caption{
 Droplet shapes on substrates with different elasticities and sliding velocities (see Movies 2-5).
 Images show droplets on stiffer ($E = 600$ kPa, panels a, b) and softer ($E = 4$ kPa, panels c, d) gels. 
 Droplet velocity increases from left to right.
 (e) Aspect ratio ($l/w$) as a function of velocity, demonstrating greater shape change on the softer substrate.
 The colored shadow represents the standard deviation across three experiments.}
 \label{fig:2}
\end{figure*}

We employ a Linnik interference imaging setup to quantify soft substrate deformation \cite{dubois2002high, xu2020viscoelastic}.
An LED light source ($\lambda = 643$ nm) illuminates the substrate from below, and reflected light from the water-substrate interface interferes with a reference beam.
Figure~\ref{fig:2} displays typical captured patterns.
{These interference patterns, when processed with phase retrieval methods, enable high-resolution analysis of substrate deformation on the order of 1 nm (details and demo code in ESI).}
Additionally, the interferometric imaging simultaneously captures droplet shapes (dark shadows, Fig.~\ref{fig:2}) and capillary positions (light shadows, Fig.~\ref{fig:2}).

\section{Experimental Results and Discussions}

\subsection{Force measurements}\label{sec:force}

Figure~\ref{fig:1}(c) shows direct measurements of the drag force $F$ experienced by a droplet moving on a soft substrate as a function of droplet velocity, $v$.
Data for 4 kPa silicone gels and 600 kPa silicone elastomers are shown as filled red squares and hollow blue circles, respectively.
Consistent with previous work on viscoelastic braking, the drag force is significantly higher on the softer substrate and increases with velocity. 
Interestingly, at higher velocities, the drag force levels off.
We note that the drag force is directly connected to the energy dissipation rate, which is $P = Fv$, shown in {Fig.~S3} in ESI.

{This dissipation can arise either from the solid substrate, the contact line, or the liquid droplet.
The capillary number, $\mathrm{Ca} = \mu v / \gamma$, where the viscosity $\mu \approx 1 \times 10^{-3}$ Pa s, characterizes the relative importance of viscosity and surface tension in contact-line motion \cite{voinov1976, cox1986}.
In our experiments, Ca varies from $10^{-8}$ to $10^{-5}$, indicating that viscous flow effects are likely negligible \cite{qian2004, bonn2009, snoeijer2013}.
Furthermore, our measurements [Fig.~\ref{fig:1}(c)] show that the drag on the hard gel is much smaller than on the soft gel. This suggests that dissipation due to viscosity as well as contact-line pinning is a minor contributor to the total dissipation, particularly at low velocities.
}

{Thus, the dissipation must occur in the substrate. We compare the measured drag force to a linear viscoelastic model, adapted from Ref.~\citenum{karpitschka2015droplets}, which originally described an isotropically spreading droplet on a thin, deformable substrate.
We approximate the experimental system as a moving circular contact line with a diameter matching that of the resting droplet.
We estimate the drag force per length at each point on the contact line by decomposing the motion onto vectors perpendicular and parallel to the contact line.  We assume the component of motion along the contact line generates no drag and the component perpendicular to the contact line generates the same resistance as an isotropically expanding droplet. 
The total drag force is then obtained by integrating over the entire circular contact line.
Substituting the wetting and rheological properties of our softer gel into the model yields a theoretical prediction, shown as the dashed red line in Fig.~\ref{fig:1}(c).
Notably, this prediction overestimates the experimentally measured drag.
Further details of the model and calculations are available in the ESI.
In the next Section, we will demonstrate that a part of this overestimation is related to the change in the droplet shape.}

\subsection{Droplet shapes}\label{sec:shape}

While the simple model in Section~\ref{sec:force} conveniently assumed that the droplet shape was unchanged by motion, Refs.~\citenum{podgorski2001corners, oleron2024morphology} emphasize that droplets can undergo significant shape changes during sliding.
Our experiments confirm this, especially at higher velocities on softer substrates.
Typical droplet shapes captured by interferometric imaging during sliding from left to right are shown in Fig.~\ref{fig:2} (dark shadows represent the droplets).

\begin{figure*}[ht]
 \centering
 \includegraphics[width=1\textwidth]{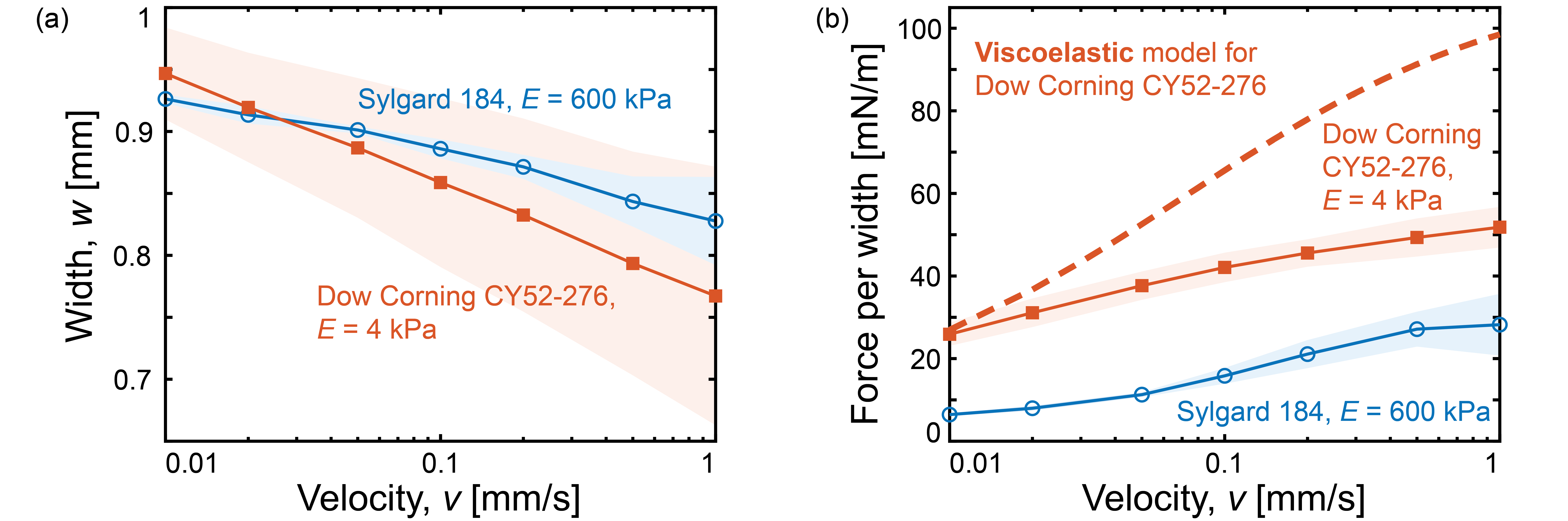}
 \caption{{The measured (a) width $w$ and (b) force per unit width of the moving droplet as a function of velocity $v$.
 The dashed line represents the prediction of a 2D, linear, viscoelastic model for a moving, circular contact line.}
 The colored shadow represents the standard deviation across three experiments.}
 \label{fig:3}
\end{figure*}

At lower velocities, droplets on both softer and stiffer gels maintain a nearly circular contact line [Figs.~\ref{fig:2}(a, c)].
However, at higher velocities, the contact line elongates in the direction of motion and narrows in the transverse direction [Figs.~\ref{fig:2}(b, d)].
This reshaping is particularly pronounced on softer gels, where the droplets take on a bullet-like form:
The contact line on the side of the droplet straightens,  parallel to the direction of motion.
We quantify this shape-change by measuring the droplet's length, $l$, and width, $w$ [Fig.~\ref{fig:2}(e)].
The aspect ratio, $l/w$, increases with velocity on both gels, with a larger shape change on the softer gel.
This increase in the aspect ratio is mainly driven by the change in the droplet width, as the change in the length of the droplet is minor [measurements in {Fig.~\ref{fig:3}(a)} and Fig.~S4 in ESI].
{Note that the contact line is expected to remain circular  in the limit of low speeds, as the net dissipative force vanishes \cite{karpitschka2015droplets}.
However, this regime has not been reached in our experiments.}

\begin{figure*}[ht]
 \centering
 \includegraphics[width=0.9\textwidth]{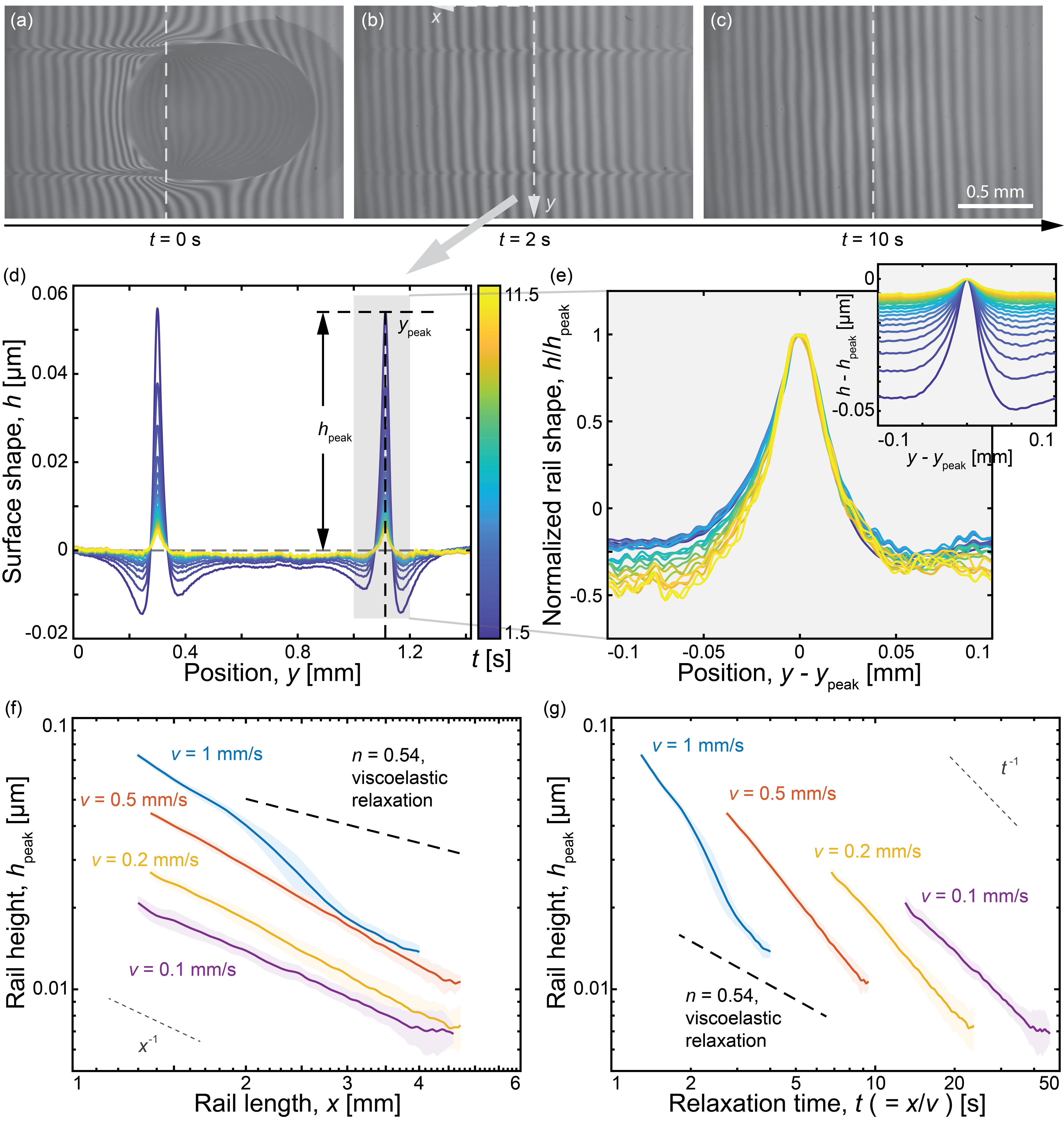}
 \caption{The relaxation of the elastocapillary rails.
 {The substrate is the softer gel ($E = 4$ kPa), and the velocity of the sliding droplet is $v = 1$ mm/s}.
 (a-c) Interferometric images showing the time evolution of the deformed substrate: $t =$ 0 s (a), 2 s (b), and 10 s (c).
 The corresponding movie is Movie 6. 
 (d) The relaxation of the substrate ($h$) along the vertical dashed line in (a-c).
 (e) Zoom-in profile of the rails near their tips.
 The rail shape is normalized by its peak height, $h_\mathrm{peak}$.
 The inset shows the rail profile (height difference from the peak, $h-h_\mathrm{peak}$) as a function of $y - y_\mathrm{peak}$.
 (f) Relaxation of the rail height $h_\mathrm{peak}$ as a function of the length along the rail, $x$.
 (g) Relaxation of the rail height $h_\mathrm{peak}$ over time, $t$.
 {The colored shadow represents the standard deviation, calculated from 737 measurements taken at different positions along the rail within the interferometry image sequences [{\it i.e.}, different x as shown in (b)] from a single experiment.
 Further details are provided in Fig.~S6 in the ESI.}
 }
\label{fig:4}
\end{figure*}

Despite confinement in the narrow gap between the capillary and the substrate, we observe a decrease in droplet width with increasing velocity, {shown in Fig.~\ref{fig:3}(a)}.
This suggests that using dynamic droplet width, rather than its size at rest, is a more accurate way to characterize the size of the contact line involved in energy dissipation.
Accordingly, we calculate the drag per unit width by dividing the measured force by twice the droplet width, Fig.~\ref{fig:3}(b).
By considering the droplet shape change, the experimental data moves closer {(but certainly not collapsing)} to the 2D linear viscoelastic model prediction [dashed red line in Fig.~\ref{fig:3}(b)], particularly at low velocities.

{The model still overestimates dissipation at high velocities, which is likely due to several simplifications.
The model assumes a linear viscoelastic response, while nonlinear effects may become significant at large deformations \cite{andreotti2020statics}.
Additionally, the model integrates contributions from independent points on the circular contact line, treating the contact line as locally straight and neglecting the potential influence of its overall curvature.
We also note that we treat the surface tensions of the solid substrate and liquid as constants \cite{xu2020viscoelastic}, while it is known that the surface tension of an aqueous liquid can decrease in contact with a silicone gel \cite{hourlier2017role}, and the surface tension of a soft solid can vary significantly under deformation \cite{xu2017, style2017elastocapillarity, bain2021}.}

\subsection{Elastocapillary Rails}

At high speeds, droplets not only assume a bullet-like morphology as previously described, but they also leave behind a striking pair of parallel rails.
These structures are clearly visible in our interferometric images {on softer gels} [Figs.~\ref{fig:2}(c, e), Fig.~\ref{fig:4}(a), and Movies 5 and 6)].
{Such rails are not observed on the stiffer elastomer.}
They appear as a pair of parallel deformations to interferometric fringes following the direction of the droplet motion, and connecting to its sides.
These fringes indicate surface topography, which we quantify using phase retrieval methods described in Refs.~\citenum{takeda1982fourier, herraez2002fast, kasim2017} (details in ESI).
This method achieves a remarkably high out-of-plane resolution on the order of 1 nm.
The resulting profile, shown in Fig.~\ref{fig:4}(d), reveals a pair of parallel ridges we term `elastocapillary rails.'
These rails can reach up to $O(1)$ mm in length. They are tallest near the droplet, with a height of $O(100)$ nm from the rear of the droplet.  Interferometry does not allow measurements closer to the contact line because steep profiles prevent resolution of the fringes.

The shape of the elastocapillary rails shares some features with typical wetting ridges, in the absence of a contact line. 
Analyzing the deformed surface shape along a vertical line in the interferometric images [Figs.~\ref{fig:4}(a-c)] reveals their cross-sectional profile [Fig.~\ref{fig:4}(d)].
They have a width of about {$50$ $\mu$m}, compared to the elastocapillary length of {$\gamma / E \approx 10$ $\mu$m}, and `dimples' on either side that are known to appear when the substrate thickness becomes comparable to the elastocapillary length \cite{style2013universal}. 
In contrast to wetting ridges, however, the tips of the rails are rounded \cite{berman2019singular, xu2020viscoelastic} rather than triangular \cite{style2013universal}.
Notably, the rails extend beyond the immediate vicinity of the contact line and detach from the droplets.
Finally, the shape of the tips of these wetting ridges evolves over time in a self-similar fashion.
If we shift the tips of the rails at different times, we see that their shapes do not overlap [inset in Fig.~\ref{fig:4}(e)].
Instead, the slopes along the sides and the curvatures at the tip decrease over time or distance from the droplet. 
However, scaling the rail profile by its peak height collapses the profiles at different locations/times [Fig.~\ref{fig:4}(e)]. Self-similar evolution is consistent with the viscoelastic relaxation model \cite{karpitschka2015droplets}.

We quantify the rail height ($h_\mathrm{peak}$) [Fig.~\ref{fig:4}(d)]  as a function of distance from the droplet, $x$.
Rails are more prominent at higher droplet velocities, as shown in Fig.~\ref{fig:4}(f).
Conversely, Fig.~\ref{fig:4}(g) displays the relaxation of the height as a function of time since it was shed from the sliding droplet. 
Here, lower droplet velocities result in longer rail persistence.

\section{Discussion and Conclusions}

Our interferometric platform enabled simultaneous investigation of multiscale structure and dynamics of droplets sliding on a deformable substrate.
At higher droplet velocities, droplets adopt a distinct bullet-like shape, the drag force levels off, and prominent parallel elastocapillary rails are shed from the droplet.

{In contrast to our previous work with similar-sized droplets sliding under gravity \cite{smith2021droplets}, where the bullet-like shape indicative of elastocapillary rails was not observed, the droplets in our current study move at speeds approximately 100 times faster. This suggests that elastocapillary rails for millimeter-sized droplets are likely limited to strongly driven systems, such as the shear geometry employed here or potentially in a spin coater.}

Feedback between the microscopic structure of the elastocapillary ridge and the overall shape of the droplet plays an essential role in these phenomena.
Sliding along a wetting ridge creates much less drag than surfing across it.
This favors the narrowing of the droplet, creating longer droplets.
At high speeds, the length of the parallel side contact lines of the bullet-like droplets is similar to the droplet length, $l$.

{We hypothesize that elastocapillary rails are shed at high speeds because points along the sides of the droplet are pulled upward for about $100\times$ longer than points along its front or back. 
Specifically, points along the sides are pulled upward by the liquid-vapor interface for a time on the order of $l/v$, from roughly 1 to 100 s. Points at the front and back are only pulled upward for a time on the order of $\gamma/Ev$, from roughly 0.01 to 1 s.
The relaxation data in Fig.~\ref{fig:4}(g) supports this hypothesis, as rails formed at lower speeds ({\it i.e.}, longer contact times) are taller.}

{The growth and relaxation of the elastocapillary rails are dependent on the mechanism of dissipation in the substrate. This may be viscoelastic or poroelastic in origin.  
Previous work \cite{zhao2018growth, xu2020viscoelastic} has shown that poroelasticity can lead to ridge decay times comparable to the contact time between the droplet and substrate.
While the viscoelastic timescale ($\tau \sim 10^{-1}$ s) governs the short-time relaxation of a wetting ridge, the poroelastic timescale ($\tau_\mathrm{p}$) governs the long-time relaxation, scaling as $w \gamma /(E D^*) \sim 1$ s, where $D^* \sim 10^{-8} ~\mathrm{m^2/s}$ is the poroelastic diffusion coefficient \cite{xu2020viscoelastic}.
Interestingly, the time for which points along the sides are pulled upward by the liquid-vapor interface (roughly 1 to 100 s) is comparable to this poroelastic timescale, while the time for which points at the front and back are pulled upward (roughly 0.01 to 1 s) is closer to the viscoelastic relaxation time.
Consequently, due to poroelasticity, elastocapillary ridges formed along the direction of motion would be more persistent than those at the leading and trailing edges of the droplet, enhancing the formation of sheddable elastocapillary rails.}

Further evidence for poroelastic contributions can be found in the rail height as a function of time, shown in Fig.~\ref{fig:4}(g). 
A purely viscoelastic model would predict $h \propto t^{-n}$, where $n = 0.54$ based on the rheology of our gel [dashed line in Figs.~\ref{fig:4}(f, g)]. 
However, we observe a steeper power-law relaxation, consistent with Ref.~\citenum{xu2020viscoelastic}.

{The poroelastic argument suggests a potential effect of droplet size on the rails: for larger droplets, the longer pull-up time could lead to slower rail decay \cite{xu2020viscoelastic}. Investigating this size dependence is an interesting topic for future research. Additionally, the influence of poroelasticity on dissipation remains an intriguing open question to explore in future studies.}

Our results underscore the need for scale-crossing models that link contact line dissipation to overall droplet shape \cite{le2005shape}, incorporating elements of both viscoelasticity \cite{karpitschka2015droplets} and poroelasticity \cite{zhao2018growth}.

\section*{Author Contributions}
N.X. and E.R.D. conceptualized the research.
N.X. and L.A.W. designed the experimental setup.
N.X. performed the experiments and processed the data.
N.X., L.A.W., R.W.S., and E.R.D. discussed the results and wrote the paper.

\section*{Conflicts of Interest}
There are no conflicts to declare.

\end{document}